\documentclass[aps,preprint]{revtex4}

 \usepackage{epsfig,rotating,minitoc}
\usepackage{amsmath,amssymb,amsthm}
\usepackage{graphicx}
\usepackage{psfrag}
\usepackage{bm}
\usepackage[dvips]{color}

\begin{document}

\title{\bf Spherical Density Functional Theory}

\author{ \'A. Nagy$^{1}$, K. Kokko$^{2}$, J. Huhtala$^{2}$, T. Bj\"orkman$^{3}$ and L. Vitos$^{4,5,6}$}
\affiliation{$^1$ Department of Theoretical Physics, University of  
Debrecen,
 H--4002 Debrecen, Hungary}
\affiliation{$^2$ Department of Physics and Astronomy, University of
 Turku, FI-20014 Turku, Finland}
\affiliation{$^3$ Faculty of Science and Engineering, Abo Akademi University, FI-20500 Turku, Finland}
\affiliation{$^4$
Applied Materials Physics, Department of Materials Science and Engineering, Royal Institute of Technology, Stockholm SE-100 44, Sweden}
\affiliation{$^5$Institute for Solid State Physics and Optics, Wigner Research Centre for Physics, Hungarian Academy of Sciences, P.O. Box 49, H-1525 Budapest, Hungary}
\affiliation{$^6$Department of Physics and Astronomy, Division of Materials Theory, Uppsala University, Box 516, SE-75121 Uppsala, Sweden}

\date{\today}

\begin{abstract}
Recently, Theophilou (J. Chem.Phys  {\bf 149} 074104 (2018)) 
showed that a set of spherically symmetric 
densities determines uniquely the external potential in molecules and solids.
Here, spherically symmetric 
Kohn-Sham-like equations are derived. The spherical densities can be 
expressed with radial wave functions. 
Expression for the total energy is also presented.

\end{abstract}

\maketitle

\section{Introduction}

Recently, Theophilou \cite{theo18} proposed a novel form of the density 
functional theory. He showed that a set of the spherical averages of the 
density around the nuclei determines uniquely
 the external potential in case of atoms, molecules or solids.
Afterwords \cite{nagy18}, an alternative proof to this theorem was put forward
and  the theory was generalized via constrained search. 
 Euler equations have also been derived. There are as many  Euler equations as
the number of nuclei. It has also been
shown  \cite{nagy20}
that any of the spherically symmetric 
densities obeys a Schr\"odinger-like differential equation which is
equivalent to the Euler equation of this density. The exact effective potential is
presented explicitly in terms of wave-function expectation values.

The
 Kohn-Sham equations have been obtained via constrained search \cite{nagy18}. 
The Kohn-Sham potential is proved to be a functional of the set of 
spherically symmetric densities. While the  Euler equations are
spherically symmetric, the Kohn-Sham equations do not have 
spherical symmetry.

It is natural to ask the question whether it is possible to derive
spherically symmetric Kohn-Sham equations that generate the 
spherically symmetric densities. Here we show that the answer is
affirmative. We argue that the reduction to spherically symmetric
equations leads to  an enormous simplification provided that 
an accurate enough approximation to the unknown functional is available.

The following arrangement is adapted. The main results of the papers
 \cite{theo18} and \cite{nagy18} are summarized in the
following section. The spherical Kohn-Sham-like equations are
derived in Section 3.
 The last section is dedicated to discussion.
The classical electron-electron potential and energy are compared with
the usual ones in the appendix.

\section{DFT with spherically averaged densities}

The Hamiltonian has the form
\begin{eqnarray}
\label{kens1}
\hat H = \hat{T} +  \hat{V}_{ee} +  \hat{V} ,
\end{eqnarray}
 where $\hat{T}$ and $\hat{V}_{ee}$ are the kinetic energy and the 
electron-electron energy operators and
\begin{eqnarray}
\label{kens1r}
{\hat V} = \sum_{i=1}^N v({\bf r}_i) 
\end{eqnarray} 
\begin{eqnarray}
\label{dv4}
v({\bf r}) = -\sum_{\beta=1}^M \frac{Z_{\beta}}{|{\bf r}-{\bf R}_{\beta}|}  
\end{eqnarray}
is the external potential. $N$, $M$, $Z_{\beta}$ and $R_{\beta}$ denote
 the number of electrons, the number of nuclei, the atomic number and the position
 vector  of the nuclei, respectively.

Theophilou constructed \cite{theo18}
the spherical 
average of the electron density $\varrho({\bf r})$  with respect to the 
nucleus $\beta$ 
\begin{eqnarray}
\label{kato2n}
{\bar \varrho}_{\beta}(r_{\beta})   = \frac{1}{4 \pi} \int_{\Omega_{\beta}} \varrho({\bf r}) d\Omega_{\beta} \;,
\end{eqnarray}
where  $\Omega_{\beta}$ denotes the angles.
Let the symbol $\{ {\bar \varrho}\}$ stand for 
the set of spherically symmetric densities 
 ${\bar \varrho}_1, {\bar \varrho}_2, ..., {\bar \varrho}_M$. 
It has been proved that this set determines the external potential.
A functional $Q$ of the set has been defined   \cite{nagy18}
using the constrained search of Levy \cite{ml79} and Lieb \cite{lieb} as
\begin{eqnarray}
\label{coulks9}
 Q[\{ {\bar \varrho}\}]= 
\mathop{{\rm min}}_{\Psi \to \{ {\bar \varrho} \} }
  \langle \Psi  | {\hat T} +{\hat V}_{ee} | \Psi \rangle \;,
\end{eqnarray}
where the minimum is searched  with the
constraint that each wave function  generates the set  $\{ {\bar \varrho}\}$.
It has also been proved that 
there exists a one-to-one map between the density and the 
set  of sperically symmetric densities $\{ {\bar \varrho}\}$.
If  $Q[\{ {\bar \varrho}\}]$
 is functionally differentiable we can arrive at
 the Euler equations
\begin{eqnarray}
\label{coulks20}
v_{\beta}(r_{\beta}) = - \frac{\delta  Q}{\delta {\bar \varrho}_{\beta}}  \mbox{;} \quad \beta = 1, ..., M
\end{eqnarray}
up to a constant. 
There are as many Euler equations as the number of the nuclei.

The non-interacting Hamiltonian ${\hat H}^{0}$ has the form
\begin{eqnarray}
\label{coulks15d}
 {\hat H}^{0} =  {\hat T} + \sum_{i=1}^N w({\bf r}_i) ,
\end{eqnarray}
where
\begin{eqnarray}
\label{dv4ks}
w({\bf r}) = \sum_{\beta=1}^M w_{\beta}(r_{\beta})  .
\end{eqnarray}
The potential $w$ is determined by the condition  that
 the set $\{ {\bar \varrho}\}$  calculated with  the
non-interacting wave function $\Phi$ be the same as the original set \cite{nagy18}.

In the absence of degeneracy  $\Phi$ can 
be given by one-particle functions and the Kohn-Sham equations takes the form 
\begin{eqnarray}
\label{coulks43}
\left [ -\frac12 \nabla^2 + w({\bf r})\right ]\phi_i = \varepsilon_i \phi_i  , 
\end{eqnarray}
where the
density $\varrho$  is 
\begin{eqnarray}
\label{coulks45}
\varrho = \sum_{i=1}^N |\phi_i|^2  .
\end{eqnarray}

The non-interacting kinetic energy  functional has been defined as \cite{nagy18}
\begin{eqnarray}
\label{coulks25}
 K[\{ {\bar \varrho}\}]= 
\mathop{{\rm min}}_{\Phi \to \{ {\bar \varrho}\}}
  \langle \Phi  | {\hat T} | \Phi \rangle \;.
\end{eqnarray} 
The minimization is done over all  wave functions $\Phi$ that
yield the same set of spherically symmetric  densities as 
the true interacting system. Writing 
the total energy of the non-interacting system in the form
\begin{eqnarray}
\label{c22pl}
E_s[\{ {\bar \varrho}\}] = K[\{ {\bar \varrho}\}] +  4 \pi \sum^{M}_{\beta=1}\int {\bar  \varrho}_{\beta}(r_{\beta}) w_{\beta}(r_{\beta}) r_{\beta}^2 d r_{\beta} 
\end{eqnarray}
 the  non-interacting Euler equations can be gained
\begin{eqnarray}
\label{coulks20v}
w_{\beta}(r_{\beta}) = - \frac{\delta  K}{\delta {\bar \varrho}_{\beta}}  \mbox{;} \quad \beta = 1, ..., M
\end{eqnarray}
up to a constant.

The difference of $Q[\{ {\bar \varrho}\}]$ and $K[\{ {\bar \varrho}\}]$ has
been defined as the Hartree and exchange-correlation 
functional $E_{Hxc}[\{ {\bar \varrho}\}]$ \cite{nagy18}
\begin{eqnarray}
\label{coulks2v5}
E_{Hxc}[\{ {\bar \varrho}\}] = Q[\{ {\bar \varrho}\}]- K[\{ {\bar \varrho}\}] .
\end{eqnarray}
Comparison of the Euler equations  (\ref{coulks20}) and (\ref{coulks20v})
yields
\begin{eqnarray}
\label{coulks20u}
w_{\beta}(r_{\beta}) = v_{\beta}(r_{\beta}) + v_{Hxc,\beta}(r_{\beta}) \mbox{;} \quad \beta = 1, ..., M  .
\end{eqnarray}
The last term in Eq. (\ref{coulks20u})
\begin{eqnarray}
\label{coulks20g}
 v_{Hxc,\beta}(r_{\beta}) = \frac{\delta E_{Hxc} }{\delta {\bar \varrho}_{\beta}}
 \mbox{;} \quad \beta = 1, ..., M  
\end{eqnarray}
stands for the Hartree and  exchange-correlation potential. Now,
 we can compute the Kohn-Sham potential $w$ in  Eq. (\ref{dv4ks}) using
$w_{\beta}(r_{\beta})$ in (\ref{coulks20u}) \cite{nagy18}.

\section{Spherical Kohn-Sham-like equations}

Define the  functional 
\begin{eqnarray}
\label{coulks25b}
 K_{\alpha}[\{ {\bar \varrho}\}]= 
\mathop{{\rm min}}_{\Phi \to {\bar \varrho}_{\alpha}}
  \langle \Phi  | {\hat T} | \Phi \rangle \;.
\end{eqnarray} 
The minimization is done over all  wave functions $\Phi$ that
yield the spherically averaged density ${\bar \varrho}_{\alpha}$.
That is, we minimize the  functional 
\begin{eqnarray}
\label{c21bb}
 \langle \Phi  | {\hat T} + {\hat U}_{\alpha} | \Phi \rangle ,
\end{eqnarray}
where
\begin{eqnarray}
\label{dv4ut}
 U_{\alpha}({\bf r}_1,...,{\bf r}_N) = \sum_{i=1}^N u_{\alpha}(r^i_{\alpha})
\end{eqnarray}
and
\begin{eqnarray}
\label{dv4urt}
r^i_{\alpha} = |{\bf r}_i -{\bf R}_{\alpha}|  .
\end{eqnarray}
Expression (\ref{c21bb}) can be rewritten as
\begin{eqnarray}
\label{c22plb}
 \langle \Phi  | {\hat T} | \Phi \rangle  +  4 \pi \int {\bar  \varrho}_{\alpha}(r_{\alpha}) u_{\alpha}(r_{\alpha}) r_{\alpha}^2 d r_{\alpha} .
\end{eqnarray}
The minimization leads to the   Kohn-Sham equations 
\begin{eqnarray}
\label{coulks43b}
\left [ -\frac12 \nabla^2 + u_{\alpha}(r_{\alpha})\right ]\psi^{\alpha}_i = \epsilon^{\alpha}_i \psi^{\alpha}_i  . 
\end{eqnarray}
The  Kohn-Sham potential $u_{\alpha}$ depends only on the radial distance from the  center $\alpha$. That is, we have  ``atomic'' Kohn-Sham equations.
Introducing radial wave functions  $P^{\alpha}_i$, the kinetic energy  
$K_{\alpha}$ can be rewritten as 
\begin{eqnarray}
\label{ka1}
 K_{\alpha} = - \frac12 \sum_i \int  P^{\alpha}_i \left ( \frac{d^2 P^{\alpha}_i}{d r_{\alpha}^2} - \frac{l_i ( l_i + 1 )}{r_{\alpha}^2} P^{\alpha}_i \right ) d r_{\alpha} .
\end{eqnarray} 

The variation leads to the radial equations
\begin{eqnarray}
\label{ka3}
 - \frac12  \frac{d^2 P^{\alpha}_i}{d r_{\alpha}^2} + \frac{l_i ( l_i + 1 )}{2 r_{\alpha}^2} P^{\alpha}_i + u_{\alpha}(r_{\alpha}) P^{\alpha}_i = \epsilon^{\alpha}_i  P^{\alpha}_i .
\end{eqnarray} 
The spherically averaged density ${\bar  \varrho}_{\alpha}$ can be
expressed with the radial wave functions  $P^{\alpha}_i$
\begin{eqnarray}
\label{ka4}
 4 \pi r_{\alpha}^2 {\bar  \varrho}_{\alpha}(r_{\alpha}) =  \sum_i \lambda_i [P^{\alpha}_i(r_{\alpha})]^2  .
\end{eqnarray} 
 $\lambda_i$ are the occupation numbers and
the sum goes for the orbitals. (For convenience only one subscript is applied.)   

The radial equations (\ref{ka3}) can be derived in another way.
First, we have to write the total energy of the original interacting system
with the  spherically averaged densities:
\begin{eqnarray}
\label{ka5}
E = {\tilde K} +  {\tilde E}_{Hxc} -  4 \pi \sum^{M}_{\alpha=1}\int {\bar  \varrho}_{\alpha}(r_{\alpha}) \frac{Z_{\alpha}}{r_{\alpha}} r_{\alpha}^2 d r_{\alpha} .
\end{eqnarray} 
Define $Z$ as the sum of the atomic numbers
\begin{eqnarray}
\label{ka7}
Z = \sum^{M}_{\alpha=1} Z_{\alpha} .
\end{eqnarray} 
For a neutral system $Z = N$. It is convenient to define the non-interacting
kinetic energy ${\tilde K}$ as a sum of spherical terms $K_{\alpha}$
(Eq. (\ref{ka1}).  
However, as all these terms are constructed for all the $N$ electrons, we
have to multiply them with the factor   $Z_{\alpha}/Z$.
\begin{eqnarray}
\label{ka9}
{\tilde K}  = \sum^{M}_{\alpha=1} \frac{Z_{\alpha}}{Z}K_{\alpha} .
\end{eqnarray}

Consider the partition of ${\tilde E}_{Hxc}$ 
\begin{eqnarray}
\label{ka11}
{\tilde E}_{Hxc} = {\tilde J} + {\tilde E}_{xc}  .
\end{eqnarray}
Define the ``spherical'' classical Coulomb energy \cite{theo18} as
\begin{eqnarray}
\label{r38b}
{\tilde J}  =2 \pi  \sum_{\alpha} \frac{Z_{\alpha}}{Z} \int_0^{\infty}{\bar \varrho}(r_{\alpha}) {\tilde u}_{J}^{\alpha}(r_{\alpha}) r^2_{\alpha} dr_{\alpha} , 
\end{eqnarray} 
where
\begin{eqnarray}
\label{r5}
{\tilde u}_{J}^{\alpha}(r_1) = 4 \pi \left (\frac{1}{r_1} \int_0^{r_1}{\bar \varrho}(r_2)r_2^2 dr_2 + \int^{\infty}_{r_1}{\bar \varrho}(r_2)r_2 dr_2 \right ) .
\end{eqnarray}
We emphasize that ${\tilde J}$ differs from the usual classical Coulomb energy.
Their relationship is presented in the appendix.
The exchange-correlation term  ${\tilde E}_{xc}$ is defined by Eqs.
(\ref{ka5}) and (\ref{ka11}).
Naturally, ${\tilde E}_{xc}$ is also different from the standard
 exchange-correlation energy. It can be rewritten as
\begin{eqnarray}
\label{ka15}
{\tilde E}_{xc} =  \sum_{\alpha} \frac{Z_{\alpha}}{Z} {\tilde E}_{xc} .
\end{eqnarray}
We emphasize that ${\tilde E}_{xc}$ is a functional of all 
spherically averaged densities. It cannot be written as a sum of terms
that depend on only one of the spherically averaged densities. 

The variation of the total energy with respect to the 
radial wave function  $P^{\alpha}_i$ leads to 
the radial equations
\begin{eqnarray}
\label{ka17}
\frac{Z_{\alpha}}{Z} \left [ - \frac12  \frac{d^2 P^{\alpha}_i}{d r_{\alpha}^2} + \frac{l_i ( l_i + 1 )}{2 r_{\alpha}^2} P^{\alpha}_i + \left ( {\tilde u}_J^{\alpha} + {\tilde u}_{xc}^{\alpha} \right ) P^{\alpha}_i \right ] - \frac{Z_{\alpha}}{r_{\alpha}} P^{\alpha}_i = {\tilde \epsilon}^{\alpha}_i  P^{\alpha}_i ,
\end{eqnarray} 
where
\begin{eqnarray}
\label{ka18}
{\tilde u}_{xc}^{\alpha} = \frac{\delta {\tilde E}_{xc}}{\delta {\bar \varrho}_{\alpha}} .
\end{eqnarray} 
Eq. (\ref{ka17}) can be rewritten as 
\begin{eqnarray}
\label{ka19}
- \frac12  \frac{d^2 P^{\alpha}_i}{d r_{\alpha}^2} + \frac{l_i ( l_i + 1 )}{2 r_{\alpha}^2} P^{\alpha}_i + \left ( {\tilde u}_J^{\alpha} + {\tilde u}_{xc}^{\alpha} \right ) P^{\alpha}_i  - \frac{Z}{r_{\alpha}} P^{\alpha}_i = \epsilon^{\alpha}_i  P^{\alpha}_i ,
\end{eqnarray} 
where
\begin{eqnarray}
\label{ka21}
\epsilon^{\alpha}_i = \frac{Z}{Z_{\alpha}}  {\tilde \epsilon}^{\alpha}_i .
\end{eqnarray} 
Comparing Eqs.   (\ref{ka3}) and (\ref{ka19}) we obtain
\begin{eqnarray}
\label{ka23}
u_{\alpha}(r_{\alpha}) = {\tilde u}_J^{\alpha}(r_{\alpha}) + {\tilde u}_{xc}^{\alpha}(r_{\alpha}) - \frac{Z}{r_{\alpha}}  .
\end{eqnarray} 

\section{Discussion}

As we use the set of  spherically averaged densities as basic variable we 
write the total energy   of the original interacting system
with the  spherically averaged densities. This form (Eq. (\ref{ka5})) is 
different from the usual partition of the total energy 
\begin{eqnarray}
\label{ka5b}
E = T_s +  J + {E}_{xc} -  4 \pi \sum^{M}_{\alpha=1}\int {\bar  \varrho}_{\alpha}(r_{\alpha}) \frac{Z_{\alpha}}{r_{\alpha}} r_{\alpha}^2 d r_{\alpha} ,
\end{eqnarray} 
where $T_s$, $J$ and ${E}_{xc}$ are the standard non-interacting kinetic,
the classical Coulomb repulsion and the exchange-correlation energies,
respectively.
The last (external) term is the same in Eqs.  (\ref{ka5}) and (\ref{ka5b}), the
others are different. Only their sum is the same:
\begin{eqnarray}
\label{ka5c}
{\tilde K} + {\tilde J} + {\tilde E}_{xc} = T_s +  J + {E}_{xc} 
\end{eqnarray} 

Our main result is Eq. (\ref{ka19}). Note that in all these radial equations 
the ``external'' term is the same. We first might think that a term with 
$Z_{\alpha}$ instead of $Z$ would be more appropriate. However, in that case
we would have an ``atom'' with $N$ electrons and atomic number $Z_{\alpha}$.
As $N$ is generally much larger than  $Z_{\alpha}$, we will not have a bound
 system.  So, our equation with $Z$ is correct. Certainly, we have an
``atom'' with atomic number $Z$  only in the asymptotic limit
$r_{\alpha} \to \infty$ , that is,
very far from any  nucleus. All other value of $r_{\alpha}$ the ``effect''
of the nuclei is different from  $-Z/r_{\alpha}$. This difference is included 
in the exchange-correlation potential ${\tilde u}_{xc}$. That means that 
in our case  exchange-correlation is not universal, it depends on all
atomic numbers. Of course, it is not surprising as we do not have
independent atoms, the atomic-like equations cannot be independent,
they are related via exchange-correlation. 

\section*{Appendix}

Consider the  classical Coulomb interaction potential 
\begin{eqnarray}
\label{r1}
v_{J}({\bf r}_1) = \int d{\bf r}_2 \frac{\varrho({\bf r}_2)}{r_{12}} .
\end{eqnarray}
The density can be expanded around a center ${\alpha}$
\begin{eqnarray}
\label{r2}
\varrho({\bf r}_{\alpha}) = {\bar \varrho}(r_{\alpha}) + \sum_{l>0, m}\varrho_{l m}(r_{\alpha}) Y_{l m}(\hat{r}_{\alpha}) .
\end{eqnarray}
Subsituting the expansion
\begin{eqnarray}
\label{r3}
\frac{1}{r_{12}} =  \sum_{l m}\frac{4 \pi}{2l+1}\frac{r_<^l}{r_>^{l+1}}  Y^*_{l m}(\hat{r}_1) Y_{l m}(\hat{r}_2) 
\end{eqnarray}
into $v_{J}$ we obtain
\begin{eqnarray}
\label{r4}
v_{J}({\bf r}_{\alpha}) = u_{J}^{\alpha}(r_{\alpha}) + \sum_{l>0, m}v_J^{l m}(r_{\alpha}) Y_{l m}(\hat{r}_{\alpha}),
\end{eqnarray}
where  $u_{J}^{\alpha}(r_{\alpha})$ is given by Eq. (\ref{r5}) and
\begin{eqnarray}
\label{r6}
v_{J}^{l m}(r_1) = \frac{4 \pi}{2l+1} \left (\frac{1}{r_1^{l+1}} \int_0^{r_1}\varrho_{l m}(r_2)r_2^{l+2} dr_2 + \int^{\infty}_{r_1}\varrho_{l m}(r_2)\frac{r_1^l}{r_2^{l-1}} dr_2 \right ) .
\end{eqnarray}

Using Eq. (\ref{ka7})  we can rewrite Eq. (\ref{r4}) as
\begin{eqnarray}
\label{r14b}
v_{J}({\bf r}) = \frac1Z \sum_{\alpha} Z_{\alpha} v_{J}({\bf r}_{\alpha}) = \frac1Z \sum_{\alpha} Z_{\alpha}{\tilde u}_{J}^{\alpha}(r_{\alpha}) + \frac1Z \sum_{\alpha} Z_{\alpha}\sum_{l>0, m}v_J^{l m}(r_{\alpha}) Y_{l m}(\hat{r}_{\alpha}) .
\end{eqnarray}
The first term in  Eq.
(\ref{r14b}) is the spherical Hartree potential proposed by Theophilou \cite{theo18}
\begin{eqnarray}
\label{r28}
v_{J}^{Theo}({\bf r}) = 
\sum_{\alpha} v_{J}^{Theo\alpha}(r_{\alpha})
\end{eqnarray}
\begin{eqnarray}
\label{r28c}
 v_{J}^{Theo\alpha}(r_{\alpha}) = \frac{Z_{\alpha}}{Z} {\tilde u}_{J}^{\alpha}((r_{\alpha}) .
\end{eqnarray} 
That is, the usual  Hartree potential  can be written as
\begin{eqnarray}
\label{r14bv}
v_{J}({\bf r}) = v_{J}^{Theo}({\bf r}) + {\tilde v}_{J}({\bf r})
\end{eqnarray}
with
\begin{eqnarray}
\label{r14bm}
\tilde{v}_{J}({\bf r}) =
\frac1Z \sum_{\alpha} Z_{\alpha}\sum_{l>0, m}v_J^{l m}(r) Y_{l m}(\hat{r_{\alpha}}) .
\end{eqnarray}

The classical Coulomb interaction energy is given by
\begin{eqnarray}
\label{r7}
J = \frac12 \int d{\bf r}_1 d{\bf r}_2 \frac{\varrho({\bf r}_1)\varrho({\bf r}_2)}{r_{12}} =  \frac12
\int d{\bf r}_1 \varrho({\bf r}_1) v_{J}({\bf r}_1).
\end{eqnarray}
Subsituting $v_{J}$ into $E_J$ we obtain 
\begin{eqnarray}
\label{r8}
{J} = {\tilde J} + \sum_{l>0, m}E_J^{l m},
\end{eqnarray}
where ${\tilde J}$ (Eq.  (\ref{r38b})) is the spherical Hartree energy
 defined by Theophilou \cite{theo18}
and
\begin{eqnarray}
\label{r10}
E_{J}^{l m} = 2 \pi  \int_0^{\infty}\varrho_{l m}(r)v_{J}^{l m}(r) r^2 dr .
\end{eqnarray}
$E_J$ can also be written as 
\begin{eqnarray}
\label{r38}
{J} = {\tilde J} + {\tilde E}_{J},
\end{eqnarray}
where
\begin{eqnarray}
\label{r38c}
{\tilde E}_{J} = \sum_{l>0, m} E_J^{l m}  .
\end{eqnarray}
That is, both the usual hartree potential and the energy 
are different from the spherical forms defined by Theophilou \cite{theo18}
and applied here.




\end{document}